\documentclass[conference]{IEEEtran}

\ifCLASSINFOpdf
  \usepackage[pdftex]{graphicx}
\else
\fi
\usepackage{amsmath}
\usepackage{amsfonts}
\begin{document}
%
\title{Mutual Information Analysis of Neuromorphic Coding for Distributed Wireless Spiking Neural Networks}


\author{\IEEEauthorblockN{Pietro Savazzi}
\IEEEauthorblockA{\textit{University of Pavia} \\
\textit{\& CNIT Consorzio Nazionale}\\
\textit{Interuniversitario per le} \\
\textit{Telecomunicazioni - Unità di Pavia}\\
\textit{Pavia, Italy}\\
pietro.savazzi@unipv.it}
\and
\IEEEauthorblockN{Anna Vizziello}
\IEEEauthorblockA{\textit{University of Pavia} \\
\textit{\& CNIT Consorzio Nazionale}\\
\textit{Interuniversitario per le} \\
\textit{Telecomunicazioni - Unità di Pavia}\\
\textit{Pavia, Italy}\\
anna.vizziello@unipv.it}
\and
\IEEEauthorblockN{Fabio Dell'Acqua}
\IEEEauthorblockA{\textit{University of Pavia} \\
\textit{\& CNIT Consorzio Nazionale}\\
\textit{Interuniversitario per le} \\
\textit{Telecomunicazioni - Unità di Pavia}\\
\textit{Pavia, Italy}\\
fabio.dellacqua@unipv.it}}


%


\maketitle

\begin{abstract}
Wireless spiking neural networks (WSNNs) allow energy-efficient communications, especially when considering edge intelligence and learning for both terrestrial beyond 5G/6G and space networking systems.
Recent research work has revealed that distributed wireless SNNs (DWSNNs) show good performance in terms of inference accuracy and low energy consumption of edge devices, under the constraints of limited bandwidth and spike loss probability. Following this reasoning, this technology can be promising for wireless sensor networks (WSNs) in space applications, where the energy constraint is predominant.
In this work, we focus on neuromorphic impulse radio (IR) transmission techniques for DWSNNs, quantitatively evaluating the features of different coding algorithms that can be viewed as IR modulations. Specifically, the main contribution of this work is the evaluation of information-theoretic measures that may help in quantifying performance trade-offs among existing neuromorphic coding techniques.
\end{abstract}


%
\IEEEpeerreviewmaketitle

\section{Introduction}
Wireless spiking neural networks (WSNNs) are emerging as a promising technology for energy-efficient communication in future networks \cite{Liu2024}, particularly for applications involving device-to-device (D2D) or vehicle-to-everything (V2X) communication. This technology has significant potential beyond 5G and 6G systems, where edge intelligence and learning play a crucial role \cite{Borsos2022}.

Recent studies have demonstrated the effectiveness of distributed Wireless SNNs (DWSNNs) in achieving high inference accuracy and low energy consumption for edge devices. These DWSNNs operate under the constraints of limited bandwidth and the possibility of data loss during transmission characterized by the probability of spike loss \cite{Borsos2022}.

The low energy consumption feature can be exploited in WSNs for space applications \cite{Gamba2014}\cite{Arpesi2014}, where power saving is a primary objective \cite{Furano2020}\cite{Kucik2021}, especially if considering edge learning\cite{Yan2021}.

In \cite{Chen2023}, the authors introduce a comprehensive design for a neuromorphic wireless Internet of Things (IoT) system that combines spike-based sensing, processing, and communication. The suggested system merges neuromorphic sensing and computing with impulse radio (IR) transmission, allowing energy consumption to adjust according to the monitored environment's dynamics while maintaining low-latency inference.

An innovative low-power "all-spike" approach is presented in \cite{Skatchkovsky2020}, for remote wireless inference using neuromorphic sensing, impulse radio (IR) and spiking neural networks (SNN). The suggested comprehensive neuromorphic edge architecture offers a promising platform for low-latency, efficient remote sensing, communication, and inference.

For WSNN, AI-native communications can be based on directly sending spikes, or pulses, by encoding information for radio signalling via low power impulse radio (IR) \cite{Skatchkovsky2020}. For this purpose, different neuromorphic coding, or modulation, algorithms can be used \cite{Guo2021}, and among them, in \cite{Chen2023_ICS}, phase-coded digital data are transmitted and modulated using ultra-wideband (UWB), pulse position modulation (PPM).

In this work, we discuss the topic of developing Artificial Intelligence (AI)-native IR transmissions, focusing on comparing the different neuromorphic coding techniques of SNN \cite{Guo2021} in terms of mutual information between the input and the output of the wireless channel, modeled as additive white Gaussian noise (AWGN).

In particular, the relation between mutual information and the minimum mean square error (MMSE) in Gaussian channels\cite{Verdu2005} is used to evaluate mutual information between the transmitted IR signal at the input layer and the received signal at the wireless sensor node receiver.

This information-theoretic analysis could help in evaluating the transmission effects on the different types of neuromorphic coding, comparing these results with other performance metrics such as inference accuracy and inference accuracy loss \cite{Guo2021}.

As far as the authors know, this is the first work in which such an analysis is proposed and used for this purpose. Moreover, some neuromorphic codes described in \cite{Guo2021} have never been used in IR DWSNN transmissions.

The remainder of this paper is organized as follows. Section \ref{CodingSchemes} describes the neuromorphic coding schemes that can also be seen as different ways of modulating IR signals. In section \ref{MutualInformationAnalysis}, the proposed method for evaluating neuromorphic modulation techniques is presented.

Section \ref{ResultAndDiscussion} is devoted to compare neuromorphic modulations in terms of mutual information rate under different conditions of signal-to-noise ratio (SNR). Finally, section \ref{Conclusion} includes some concluding remarks and perspective for future works.

\section{Neural Coding Schemes}
\label{CodingSchemes}

\begin{figure*}[h!]
\centering
\includegraphics[width=0.8\linewidth]{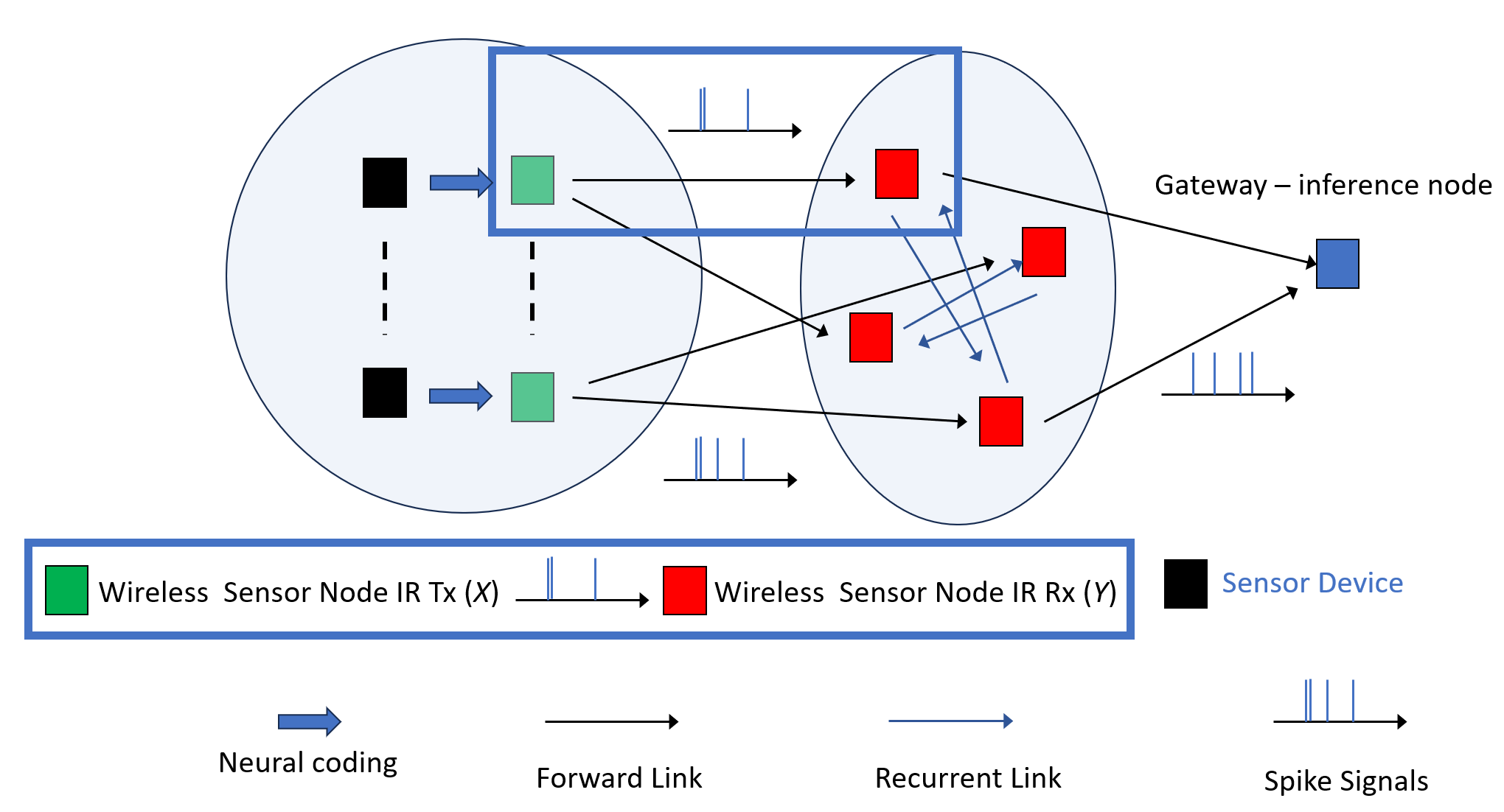}
\caption{Sysetm Model of a DWSNN.}
\label{fig_SystemModel}
\end{figure*}

In Fig. \ref{fig_SystemModel}, a DWSNN system model is shown, where the input nodes of the SNN are represented as wireless sensor devices capable of measuring and transmitting sensed signals.

Spike signals connect both hidden nodes and sensor devices, and they are radio transmitted through IR wireless modulations.

Neural coding schemes or modulations are employed to transform input signal samples into spikes, which are then sent to excitatory neurons. In the system model scheme of Fig. \ref{fig_SystemModel}, neural coding is applied to sensed signal at the input of WSN devices, to generate spike signals to be sent to the DWSNN input layer.

We consider four distinct types of neural coding schemes that were examined and compared: rate coding, time-to-first spike (TTFS) coding, phase coding, and burst coding.

The details of these schemes are detailed in the subsequent subsections. It is important to note that these coding schemes are utilized solely for encoding of input data, while output neurons remain standard neurons without encoding, for example, leaky integrated-and-fire (LIF) neuron models \cite{Guo2021}\cite{burkitt2006}.

\subsection{Rate Coding}
Rate coding (RC) is the most commonly utilized coding method in neural network models. In this approach, each input signal is treated as a firing rate and transformed into a Poisson spike train based on that rate. Input signals can be reduced by a suitable factor in order to control at the same time both the power of the modulated IR signal and its bandwidth.

\subsection{Time-to-First Spike Coding}
Time-to-first-spike (TTF) coding was identified as a method to encode information for rapid responses within a few milliseconds, such as tactile stimuli \cite{johansson2004}, utilizing initial spikes. In \cite{Park2020}, a fast and energy-efficient TTFS coding scheme was proposed, which employed an exponentially decaying dynamic threshold to transform the magnitude of the input signal into first spike patterns.

The normalized input signal is compared with a threshold generated by the following exponenatial function:
\begin{equation}
\label{TTFth}
    P_{th}(t)=\theta_0 e^{-t/\tau_{th}}
\end{equation}
A spike occurs when the amplitude of the input signal exceeds the threshold, and the input is prevented from producing additional spikes. The exact moments at which input spikes occur represents the amount of information these spikes transmit to postsynaptic neurons during the decoding phase.

The input neuron impacts the postsynaptic neuron solely at the moment it fires, and its influence does not build up over time. This streamlined approach removes the necessity for vector-matrix multiplication within the network, potentially enhancing the performance of hardware systems \cite{Guo2021}.

\subsection{Phase Coding}
According to \cite{Kim2018DeepNN}, a straightforward phase coding (PC) method is explained that transforms the amplitudes of the input signal into their binary form. The phase details are incorporated into the spikes by giving distinct weights to each bit in the burst, adhering to the typical binary representation.

The larger the input signal, the more significant spikes it generates, leading to a higher amount of transmitted information.

\subsection{Burst Coding}
Burst coding (BC) allows for rapid and efficient data transmission by emitting a cluster of spikes simultaneously. Generating multiple spikes rather than a single one can enhance the reliability of synaptic communication between neurons. Research has shown that in burst coding, information is encoded in the number of spikes (N) and the inter-spike interval (ISI) within the burst \cite{IZHIKEVICH2003161}\cite{eyherabide2009}.

In \cite{Guo2021}, they propose a simple and effective burst coding scheme that converts input signals into spike bursts with the number of spikes and ISI proportional to the signal amplitudes.

\section{Mutual Information Analysys}
\label{MutualInformationAnalysis}
In \cite{Guo2021}, the neural coding schemes described in the previous paragraph are compared by considering several performance attributes. In more detail, computational performance, i.e. inference accuracy and inference loss, is considered under different operating conditions, such as synaptic noise that could be thought of as additive white Gaussian noise (AWGN) for DWSNNs.

Taking into account arbitrarily distributed finite power input signals observed through an AWGN channel, in \cite{Verdu2005}, an interesting relation has been derived between the input-output mutual information and the MMSE, achievable by optimal estimation of the input given the output.

Specifically, the mutual information measured in nats and the MMSE adhere to the following relationship, independent of the input statistics:

\begin{equation}
    \label{MIvsMMSE}
    \frac{d}{d\mbox{snr}}I(\mbox{snr})=\frac{1}{2}\mbox{mmse(snr)}\log e
\end{equation}

where the signal-to-noise ratio (SNR) of the channel is denoted by $\mbox{snr}$, so that $\mbox{mmse}(\mbox{snr})$ and $I(\mbox{snr})$ are respectively the mmse and the mutual information considered as monotone functions of SNR, while the base of logarithm is consistent with the mutual information unit.

As explicitly stated above, we apply this analysis at the input of the DWSNN, with regard to the relationship between the output of the neuromorphic encoder, i.e. the transmitted coded IR signal $X$, and the received one $Y$, as shown in Fig. \ref{fig_SystemModel}:

\begin{equation}
\label{ChIO}
    Y=\sqrt{\mbox{snr}}X+N
\end{equation}

where $N\sim \mathcal{N}(0,\,1)$ is a standard Gaussian random noise independent of $X$. Thus, $X$ and $Y$ can be considered as input and output, respectively, of a Gaussian channel with a signal-to-noise ratio (SNR) of $\mbox{snr}$.

The MMSE estimate of the input $X$ given the output $Y$ is achieved by the conditional mean estimator:

\begin{equation}
    \label{MMSEest}
    \hat{X}(Y;\mbox{snr})=\mbox{E}\{X|Y;\mbox{snr}\}
\end{equation}

Since input-output mutual information is the well-known channel capacity under input power constraint \cite{Shannon48}

\begin{equation}
    \label{Capacity}
    I(\mbox{snr})=\frac{1}{2}\log(1+\mbox{snr})
\end{equation}

from which the MMSE can be derived as:

\begin{equation}
    \label{MSEder}
    \mbox{mmse}(\mbox{snr})=\frac{1}{1+\mbox{snr}}
\end{equation}

From equation (\ref{Capacity}-\ref{MSEder}), (\ref{MIvsMMSE}) can be derived \cite{Verdu2005}.

In this work, equations (\ref{MIvsMMSE}) and (\ref{MMSEest}) are used to calculate the derivative of the information rate between the output of the neuromorphic encoder and the received IR signal in the first layer of the DWSNN system.

In the following subsections, the system model taken into account is detailed by considering the parameters of the modulators, that is, the neuromorphic coders.

\subsection{System Model and Parameters}
In order to compare the four coding techniques described in sections \ref{CodingSchemes}, we consider as input of the coder a sensed signal represented by $N_b$ bits, i.e., quantized with $2^{N_b}$ levels.

Moreover, the four coded signals are assumed to have the same bandwidth, i.e. the same normalized impulse duration $T_i$, while the spike train duration is set to $T=N_i T_i$, where $N_i$ can be viewed as the spreading factor of the IR transmitted signal.

In the following subsections, we detail the other specific model parameters for each of the four coding schemes. It is important to note that the neural coding parameters are chosen in order to best exploit the IR modulation bandwidth. Ongoing work will also be devoted to take into account the parameters that optimize, at the same time, both the wireless transmitted spectral efficiency and inference performances.

\subsubsection{Rate Coding Parameters}
The input sample, quantized with levels of $2^{N_b}-1$ and with the amplitude normalized to 1, is considered the firing rate and, at each sampling time $T_i$, is compared to a random threshold uniformly generated in the interval $(0,1)$. If the firing rate is greater than the threshold, the output spike is equal to 1, otherwise it is equal to 0.

\subsubsection{Time-to-First-Spike Coding Parameters}
Considering equation (\ref{TTFth}), $\theta_{0}=1$, while $\tau_{th}=0.4$.

\subsubsection{Phase Coding Parameters}
The $N_b$ bits if the binary representation of the input signal amplitude generates $N_b$ spikes equally spaced over the burst interval $T$.

\subsubsection{Burst Coding Parameters}
Given the normalized to one signal amplitude of the input $A$, the maximum nuber of spikes per burst $N_{max}$, the number of spikes generated by the amplitude $A$ is given by:
\begin{equation}
    \label{NSpikesBurst}
    N_s=\lceil{AN_{max}}\rceil
\end{equation}
where $\lceil \cdot \rceil$ denotes the ceiling function.

On the other hand, The inter-spike interval (ISI) is computed by:
\begin{equation*}
  \mbox{ISI}(A)=\left\{
    \begin{aligned}
      & \lceil -(T_{max}-T_{min})A+T_{max} \rceil,\text{ }N_s > 1 \\
      & T_{max}, \text{ otherwise}
    \end{aligned}
  \right.
\end{equation*}
where the ISI is confined in the time interval $[T_{min},T_{max}]$ \cite{Guo2021}.

A larger input pixel results in a burst characterized by a shorter ISI and a higher number of spikes.

$T_{min}$ and $T_{max}$ are chosen in order to cover all the interval $T$, while ISI is determined in order of having $N_{max}=N_b$.

\section{Results and Discussion}
\label{ResultAndDiscussion}
In this section, the four neural coding algorithms are compared by evaluating the mutual information rate versus the AWGN noise variance.

\begin{figure}[h!]
\centering
\includegraphics[width=1\linewidth]{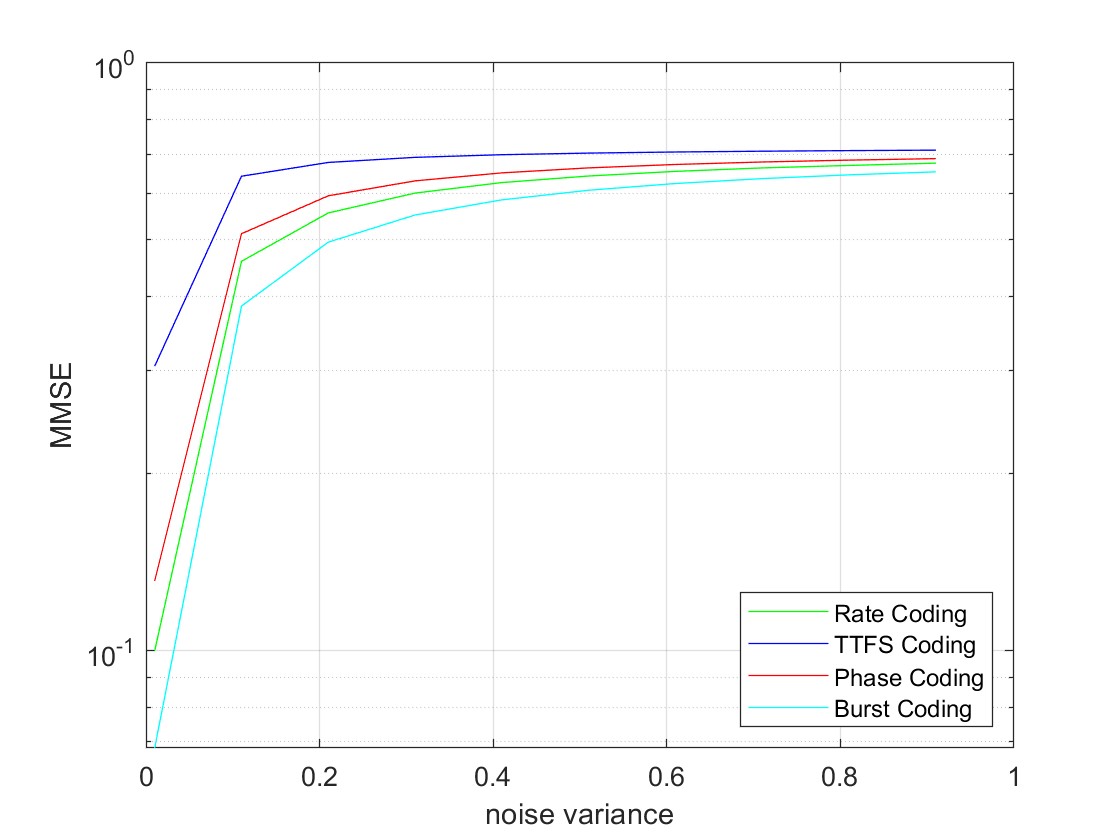}
\caption{MMSE versus noise variance, $N_b=4$, $T_i=32$.}
\label{Fig2}
\end{figure}

\begin{figure}[h!]
\centering
\includegraphics[width=1\linewidth]{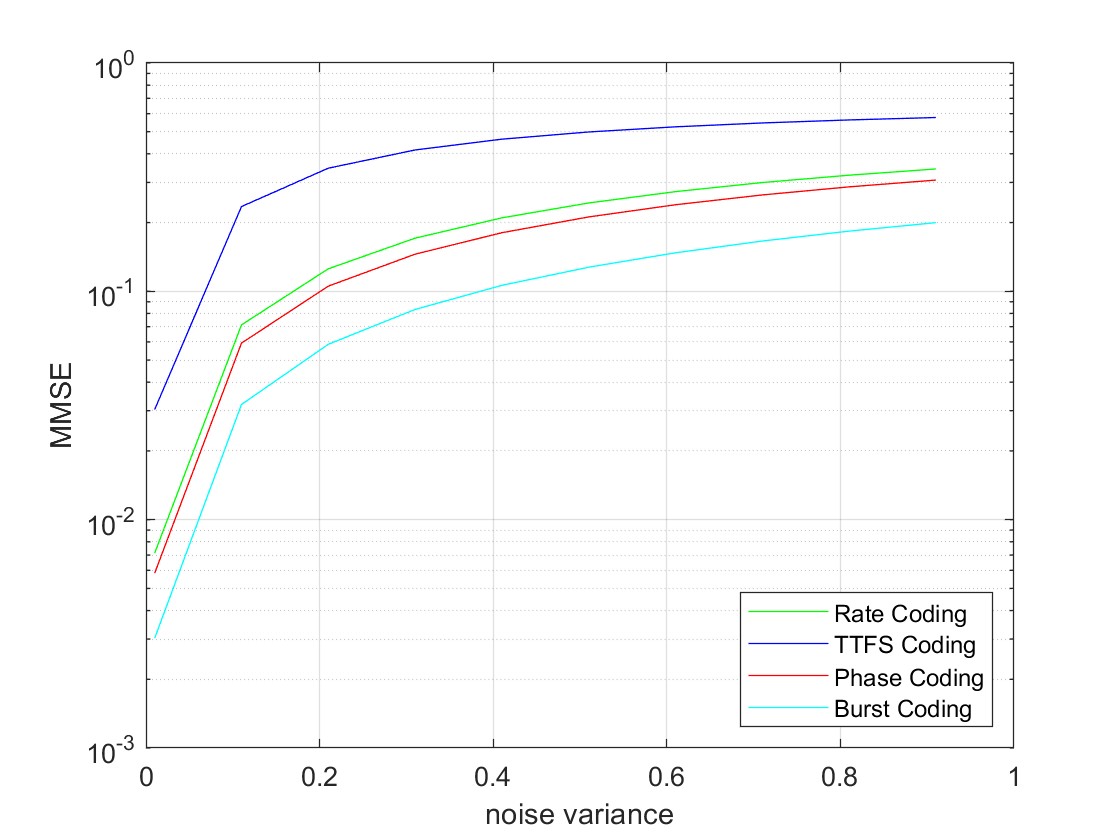}
\caption{MMSE versus noise variance, $N_b=8$, $T_i=32$.}
\label{Fig3}
\end{figure}

\begin{figure}[h!]
\centering
\includegraphics[width=1\linewidth]{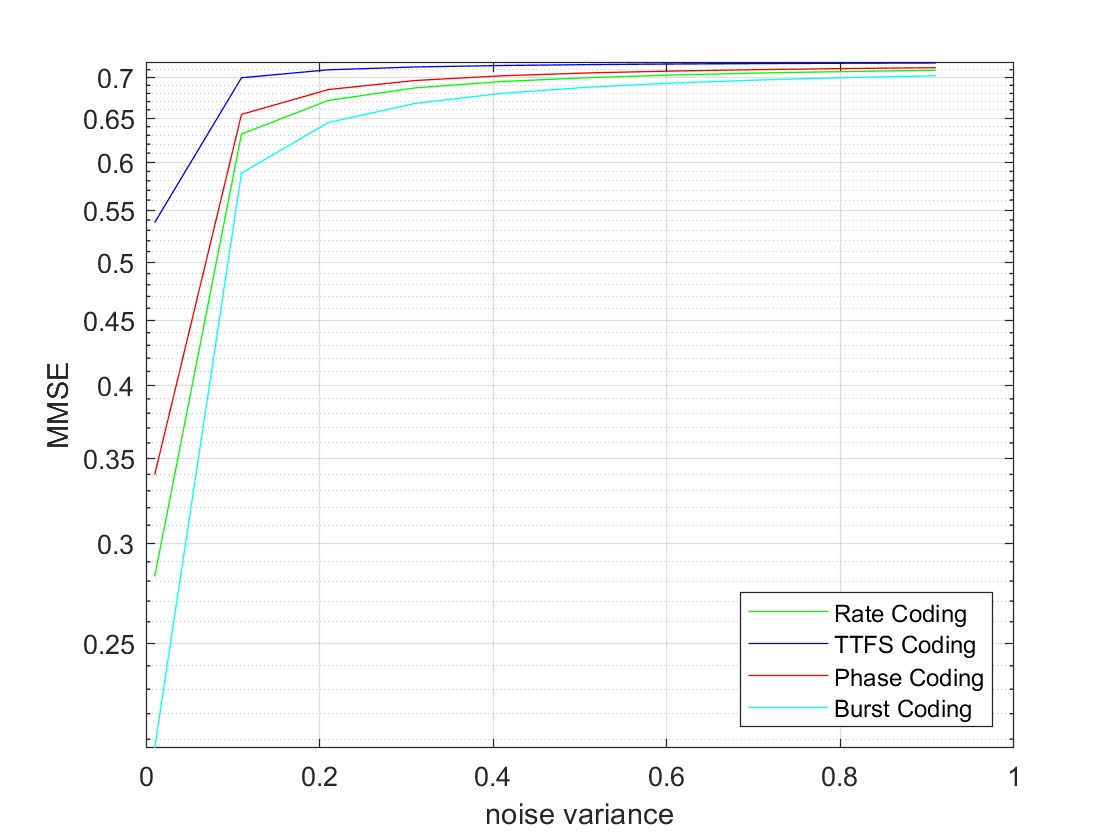}
\caption{MMSE versus noise variance, $N_b=4$, $T_i=64$.}
\label{Fig4}
\end{figure}

\begin{figure}[h!]
\centering
\includegraphics[width=1\linewidth]{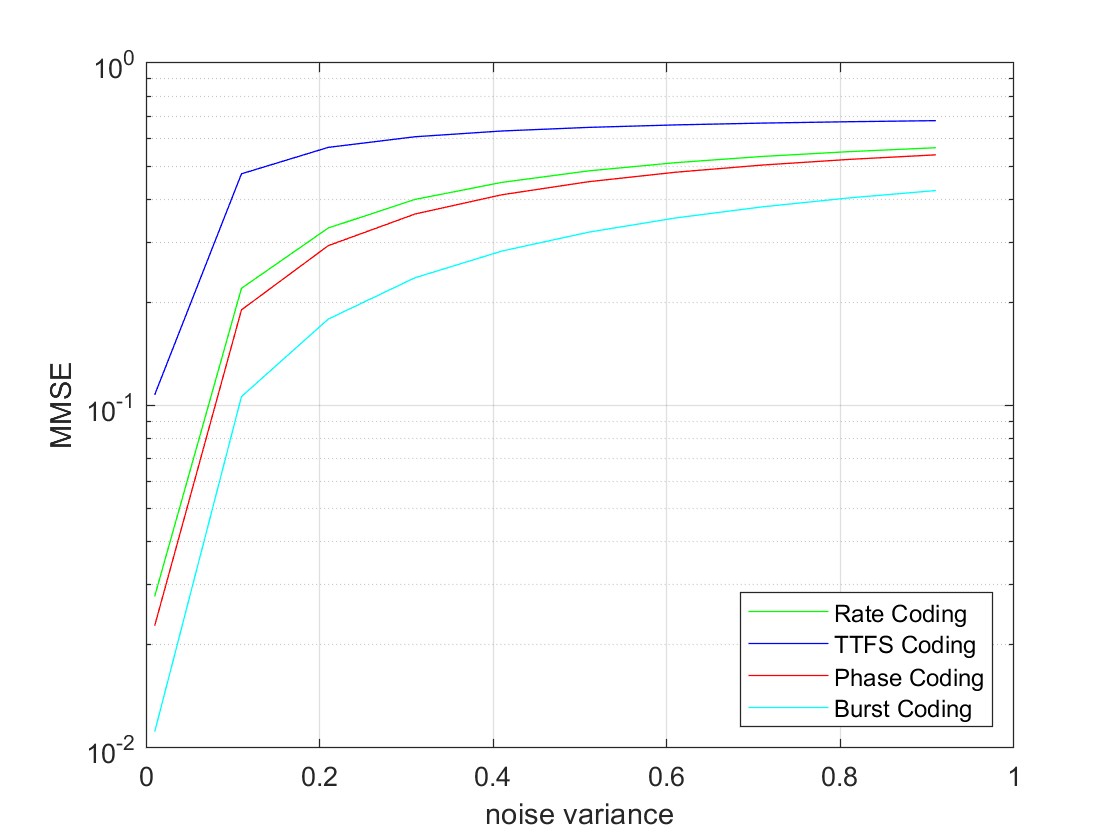}
\caption{MMSE versus noise variance, $N_b=8$, $T_i=64$.}
\label{Fig5}
\end{figure}

We consider the sensed input signal quantized with \\$N_b=4,8$ bits and an IR spreading factor $N_i$ equal to 16 and 32. The input signal samples are considered uniformly distributed along their dynamic range.

Rate and burst coding can be viewed as random coding techniques, while the phase coder output is determined by the bynary coding in a deterministic manner.

\begin{figure}[h!]
\centering
\includegraphics[width=1\linewidth]{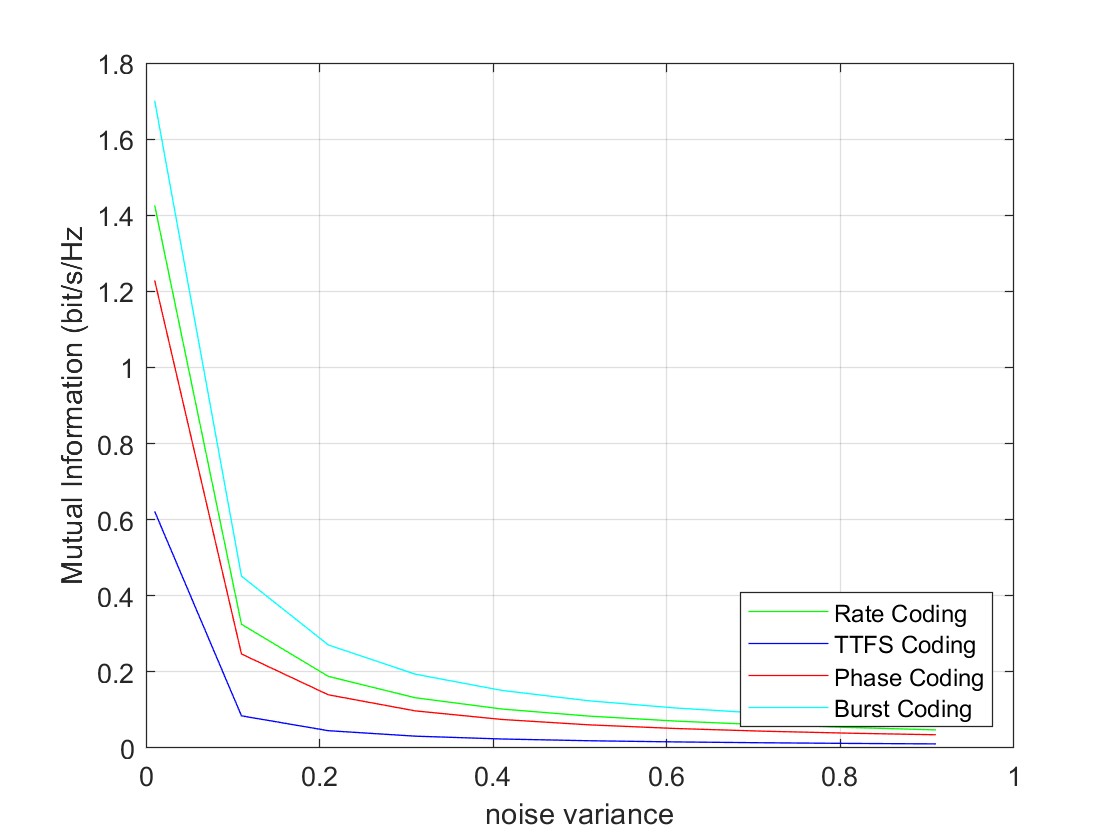}
\caption{Mutual Information versus noise variance, $N_b=4$, $T_i=32$.}
\label{Fig6}
\end{figure}

\begin{figure}[h!]
\centering
\includegraphics[width=1\linewidth]{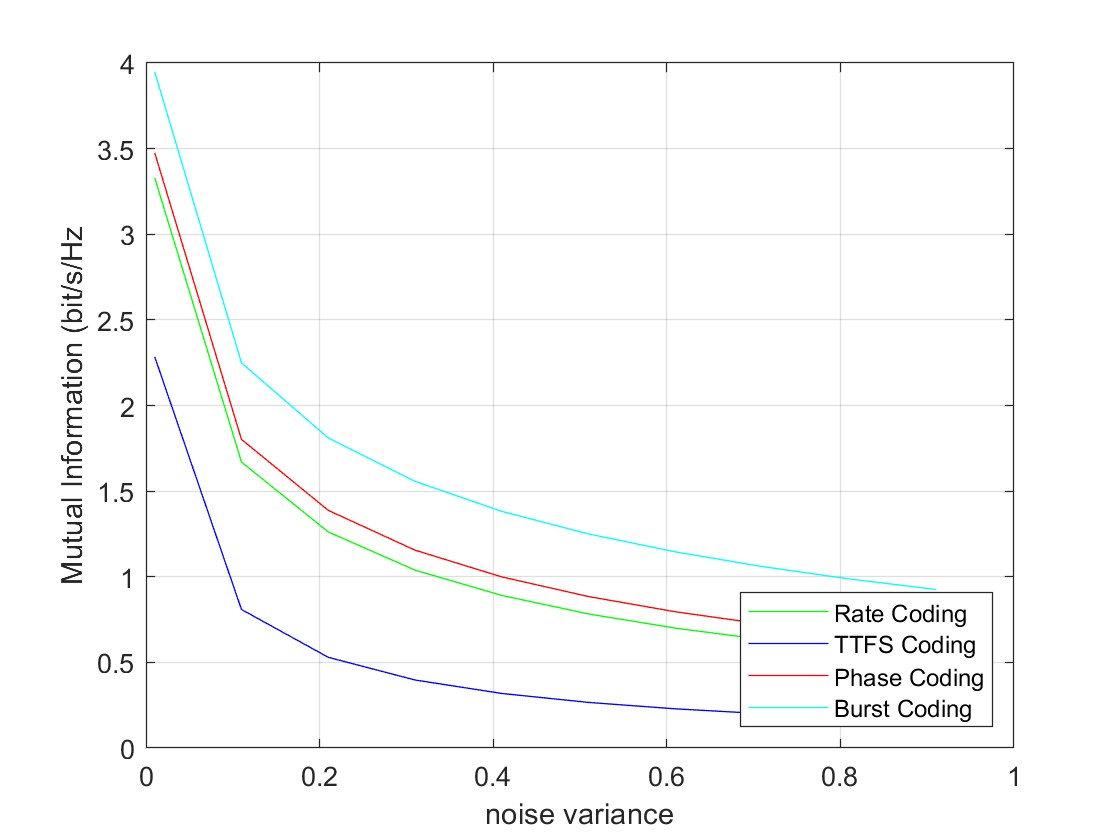}
\caption{Mutual Information versus noise variance, $N_b=8$, $T_i=32$.}
\label{Fig7}
\end{figure}

\begin{figure}[h!]
\centering
\includegraphics[width=1\linewidth]{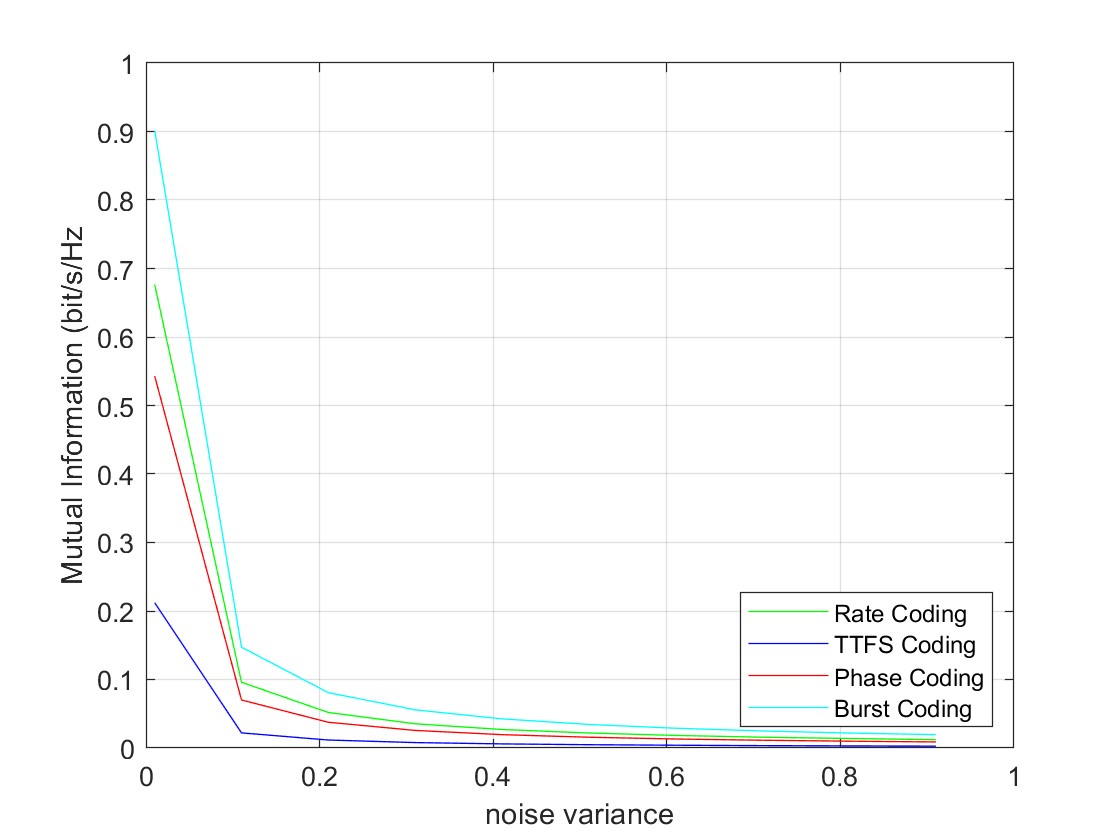}
\caption{Mutual Information versus noise variance, $N_b=4$, $T_i=64$.}
\label{Fig8}
\end{figure}

\begin{figure}[h!]
\centering
\includegraphics[width=1\linewidth]{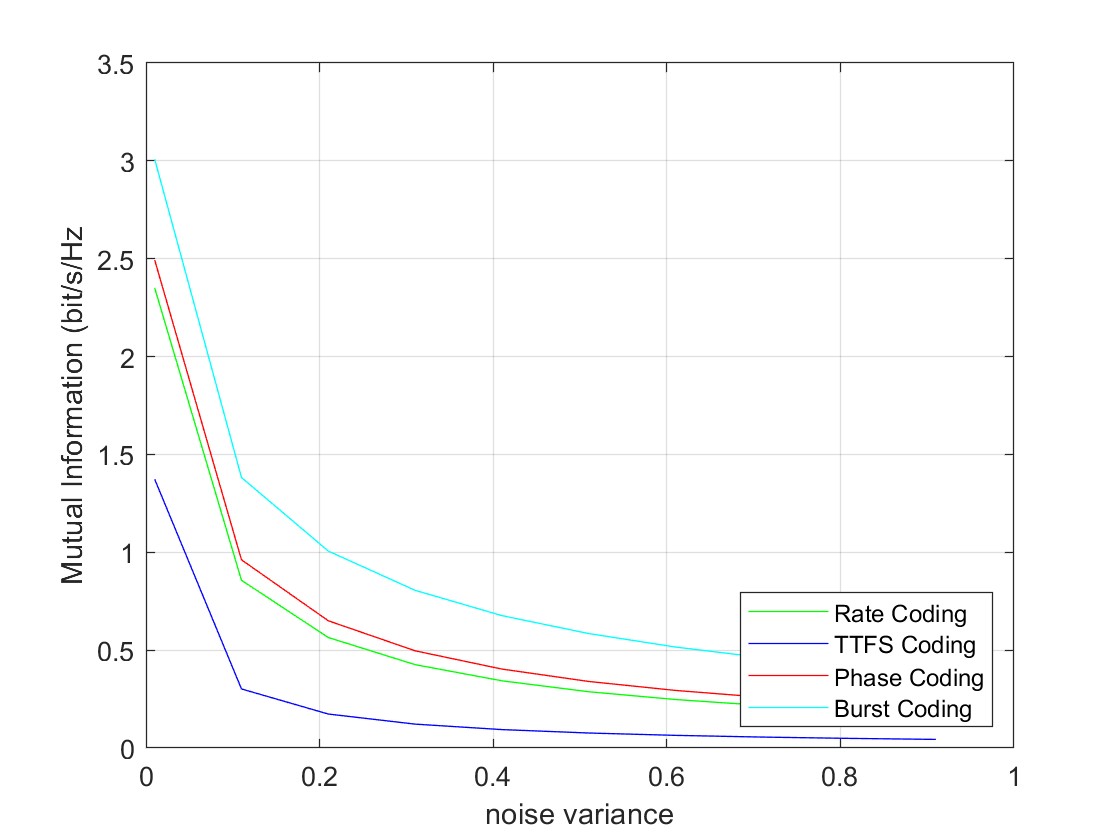}
\caption{Mutual Information versus noise variance, $N_b=8$, $T_i=64$.}
\label{Fig9}
\end{figure}

Burst coding performs better than the other in terms of MMSE of the estimated input, as shown in Figs. \ref{Fig2}-\ref{Fig5}. Obviously, similar results hold for the mutual information evaluation, as reported in Figs. \ref{Fig6}-\ref{Fig9}.

The phase and rate coder presents similar performance, while the TTFS one is slightly worse, due to the fact that the transmitter sends only one pulse per burst, like a pulse position modulation (PPM).

Even if TTFS encoding has inference accuracy performance similar to the burst one \cite{Guo2021}, BC results are much better considering the MMSE output.

\section{Conclusion}
\label{Conclusion}
As previously mentioned, Wireless Spiking Neural Networks can enable energy-efficient communications for space applications, particularly when they integrate edge intelligence and learning. Moreover, DWSNNs perform well in terms of inference accuracy and low energy consumption of edge devices, even with limited bandwidth under spike-loss probability constraints.

This work emphasizes AI-native transmission methods for DWSNN by evaluating different coding algorithms that actually represent different IR modulation techniques. Specifically, the primary outcome of this study is mutual information analysis to evaluate the performance trade-offs of standard neuromorphic coding methods.

According to the results presented, Burst Coding looks promising, especially if considering the best performance in terms of both mutual information between input and output of the wireless sensor device, and MMSE of the estimated input.

Furthermore, BC seems interesting for wireless applications, since it represents a combination of PPM ans pseudo-noise UWB modulations.

Future work will consider wireless connection protocols between all network nodes, including the hidden ones, and the possibility of embedding these techniques into current low-power standard such as LoRaWAN \cite{goldoni2018}\cite{dakic2023spiking}.

The DWSNN inference capability could also be tested on practical problems such as range and localization through RSI (Receive Strength Indicator) \cite{Savazzi2019}. Similar applications may be devoted to estimations of environment and weather parameters \cite{goldoni2018}\cite{GUERRA2024}\cite{marzi2023tillage}.

\section*{Acknowledgment}

This study was supported by the European Union under the Italian National Recovery and Resilience Plan (NRRP) of NextGenerationEU, partnership on “Telecommunications of the Future” (PE00000001 - program “RESTART”).

\bibliographystyle{IEEEtran} 
\bibliography{biblioSNN}

\end{document}